\documentclass[a4paper, twocolumn]{article}
\usepackage[utf8]{inputenc}
\usepackage{cite}
\usepackage{graphicx}
\usepackage{amssymb}
\usepackage{amsfonts}
\usepackage{amsmath}
\usepackage{bm}
\usepackage{longtable}
\usepackage{epsfig}
\usepackage{multirow}
\usepackage{authblk}%
\usepackage[paperwidth=19.5cm,paperheight=27.0cm,top=2.0cm,bottom=2.0cm,left=1.5cm,right=1.5cm]{geometry}
\usepackage[switch]{lineno}

\usepackage{fancyhdr}
\begin{document}

\title{Derivation of generalized Einstein's equations of gravitation in some non-inertial reference frames based on the theory of vacuum mechanics }
\author{Xiao-Song Wang}
\setcounter{footnote}{0}
\affil{{\normalsize Institute of Mechanical and Power Engineering, Henan Polytechnic University, Jiaozuo, Henan Province, 454000, China}}%
\date{Feb. 28th, 2024}

\twocolumn [
	\begin{@twocolumnfalse}
		\maketitle
		\begin{abstract}
			\newgeometry{left=1.0cm, right=1.0cm}%
When solving the Einstein's equations for an isolated system of masses, V. Fock introduces harmonic reference frame and obtains an unambiguous solution. Further, he concludes that there exists a harmonic reference frame which is determined uniquely apart from a Lorentz transformation if suitable supplementary conditions are imposed. It is known that wave equations keep the same form under Lorentz transformations. Thus, we speculate that Fock's special harmonic reference frames may have provided us a clue to derive the Einstein's equations in some special class of non-inertial reference frames. Following this clue, generalized Einstein's equations in some special non-inertial reference frames are derived based on the theory of vacuum mechanics. If the field is weak and the reference frame is quasi-inertial, these generalized Einstein's equations reduce to Einstein's equations. Thus, this theory may also explain all the experiments which support the theory of general relativity. There exist some differences between this theory and the theory of general relativity.

\

keywords: Einstein's equations; gravitation; general relativity; principle of equivalence; gravitational aether; vacuum mechanics.
	  \end{abstract}
	\end{@twocolumnfalse}

\



\

\

]

\section{Introduction  \label{sec 100}}
\newtheorem{assumption_my}{\bfseries Assumption}

\newtheorem{definition_my}{\bfseries Definition}

\newtheorem{lemma_my}{\bfseries Lemma}

\newtheorem{proposition_my}{\bfseries Proposition}

\newtheorem{theorem_my}{\bfseries Theorem}

\newtheorem{wcorollary_my}{\bfseries Corollary}

The Einstein's field equations of gravitation are valid in all reference frames is a fundamental assumption in the theory of general relativity \cite{MollerC1972,WeinbergS1972,MisnerC1973}. R. P. Feynman once said:"{\itshape What I cannot create, I do not understand.}" (\cite{FeynmanRP1995}, p.\  xxxii). New theories which can derive Einstein's field equations may be interesting. The reasons may be summarized as follows.

1. Many attempts to reconcile the theory of general relativity and quantum mechanics by using the techniques in quantum electrodynamics meet some mathematical difficulties (\cite{MaddoxJ1998}, p.\ 101). J. Maddox speculates that the failure of the familiar quantization procedures to cope with Einstein's equations may stem from two possible reasons. One possibility is that Einstein's equations are incomplete. The other possible reason may be that some underlying assumptions in Einstein's theory about the character of the space or time may be not suitable (\cite{MaddoxJ1998}, p.\ 101).

2. The value of the cosmological constant is a puzzle \cite{MarshD2017}. In 1917, A. Einstein thought that his equations should be revised to be (\cite{MisnerC1973}, p.\ 410)
\begin{equation}\label{Einstein 100-200}
R_{\mu\nu}-\frac{1}{2}g_{\mu\nu}R + \Lambda g_{\mu\nu} = -\kappa T^{m}_{\mu\nu},
\end{equation}
where $g_{\mu\nu}$ is the metric tensor of a Riemannian spacetime,
$R_{\mu\nu}$ is the Ricci tensor, $R\equiv g^{\mu\nu}R_{\mu\nu}$ is the scalar curvature, $g^{\mu\nu}$ is the contravariant metric tensor, $\kappa=8\pi \gamma_{N}/c^{4}$, $\gamma_{N}$ is Newton's gravitational constant, $c$ is the speed of light in vacuum, $T^{m}_{\mu\nu}$ is the energy-momentum tensor of a matter system, $\Lambda$ is the cosmological constant.

However, it seems that the cosmological constant $\Lambda$ is unnecessary when Hubble discovered the expansion of the universe. Thus, Einstein abandoned the term $\Lambda g_{\mu\nu}$ in Eqs.\ (\ref{Einstein 100-200}) and returned to his original equations (\cite{MisnerC1973}, p.\ 410). The value of the cosmological constant $\Lambda$ is also related to the energy-momentum tensor of vacuum (\cite{MisnerC1973}, p.\ 411). Theoretical interpretation of the small value of $\Lambda$ is still open \cite{MarshD2017}.

3. The problem of the existence of black hole is interesting \cite{ChoA2017}. Einstein believed that black hole cannot exist in the real world \cite{EinsteinA1939}. Recently, the Event Horizon Telescope Collaboration (EHTC) reconstructed event-horizon-scale images of the supermassive black hole candidate in the center of the giant elliptical galaxy M87 \cite{EHTC2019a}. EHTC reports that the observed image is consistent with predictions for the shadow of a Kerr black hole based on the theory of general relativity.

4. The existences and characters of dark matter and dark energy are still controversy, refers to, for instance,  \cite{HamiltonP2015,SpergelD2015,KimJ2017,BosmaA2017,GibneyE2017}.

5. The existence and characters of gravitational aether are still not clear. Sir I. Newton pointed out that his inverse-square law of gravitation did not touch on the mechanism of gravitation (\cite{WhittakerE1951}, p.\ 28;\cite{HirosigeT1968}, p.\ 91). He conjectured that gravitation may be explained based on the action of an aether pervading the space (\cite{WhittakerE1951}, p.\ 28;\cite{HirosigeT1968}, p.\ 92). In the years 1905-1916, Einstein abandoned the concepts of electromagnetic aether and gravitational aether in his theory of relativity (\cite{Kostro2000}, p.\ 27-61). However, H. A. Lorentz believed that general relativity could be reconciled with the concept of an ether at rest and wrote a letter to A. Einstein (\cite{Kostro2000}, p.\ 65). Einstein changed his view later and introduced his new concept of ether (\cite{Kostro2000}, p.\ 63-113). However, Einstein did not tell us how to derive his equations theoretically based on his new concept of the gravitational aether.

6. Whether Newton's gravitational constant $\gamma_{N}$ depends on time and space is still not clear. It is known that $\gamma_{N}$ is a constant in Newton's and Einstein's theory of gravitation. P. A. M. Dirac speculates that $\gamma_{N}$ may depend on time based on his large number hypothesis \cite{Dirac1978}. R. P. Feynman thought that if $\gamma_{N}$ decreases on time, then the earth's temperature a billion years ago was about 48$^{\circ}C$ higher than the present temperature (\cite{FeynmanRP1995}, p.\ 9).

Furthermore, there exist some other problems related to the theories of gravity, for instance, gravitational waves \cite{AbbottBP2016}, the speed of light in vacuum \cite{SamblesJR2015}, the definition of inertial system, origin of inertial force, the velocity of the propagation of gravity \cite{CornishN2017}, the velocity of individual photons \cite{GiovanniniD2015}, unified field theory, etc.

The gravitational interaction seems to differ in character from other interactions. The existing theories of gravity still face the aforementioned difficulties. Thus, it seems that new ideas about the gravitational phenomena are needed. Following Einstein \cite{EinsteinA2000}, it may be better for us to keep an open and critical mind to explore all possible theories about gravity.

In 2012, M. J. Dupre and F. J. Tipler propose an aether theory of general relativity \cite{DupreMJ2012}. Recently, generalized Einstein's equations of gravitational fields in inertial reference frames are derived based on a sink flow model of particles \cite{WangXS201908}. However, the cases of non-inertial reference frames are not discussed in Ref. \cite{WangXS201908}. The purpose of this manuscript is to propose a derivation of the Einstein's equations in some non-inertial reference frames based on the theory of vacuum mechanics \cite{WangXS200804,WangXS200810,WangXS2014PhysicsEssays,WangXS201908}.

\section{A brief introduction of the theory of vacuum mechanics \label{sec 300}}
Ancient people believe that there is a kind of continuously distributed substance which fills every corner of the space. Ancient Egyptians called this substance as nun (\cite{PinchG2002}, p.\ 172). According to ancient Egyptians, the nun is a primeval watery darkness which is continuously distributed surround the world (\cite{PinchG2002}, p.\ 172). Ancient Greeks, such as Thales and Anaximenes, believe that everything in the universe is made of a kind of fundamental substance named aether.

The thought that our universe is an infinite hierarchy which has “universes” within “universes” without end is a charming idea. Astonished by this attractive idea and inspired by the aforementioned ideas, we propose the following mechanical model of the universe.

Matter is composed of molecules. Molecules are constructed by atoms. Atoms are formed by elementary particles. Modern experiments, for instance, the Casimir effect, have shown that vacuum is not empty. Therefore, new considerations on the old concept of aether may be needed.

Thomson's analogies between electrical phenomena and elasticity helped J. C. Maxwell to establish a mechanical model of electrical phenomena \cite{WhittakerE1951}. Following J. C. Maxwell, we introduce the following assumption \cite{WangXS200804}.
\begin{assumption_my}\label{assumption 300-50}
We suppose that vacuum is filled with a kind of continuously distributed matter which may be called the $\Omega(1)$ substratum, or the electromagnetic aether.
\end{assumption_my}

The idea that all microscopic particles are sink flows in a fluidic substratum has been proposed by many researchers in the history, for instance, J. C. Maxwell (\cite{WhittakerE1951}, p.\ 243), B. Riemann (\cite{RiemannB2004}, p.\ 507), H. Poincar$\acute{e}$ (\cite{PoincareH1997}, p.\ 171), J. C. Taylor (\cite{Taylor2001}, p.\ 431-436). Thus, we suppose that electric charges are sources or sinks of the $\Omega(1)$ substratum \cite{WangXS200804}.
Maxwell's equations in vacuum are derived by methods of continuum mechanics based on a mechanical model of vacuum and a source or sink flow model of electric charges \cite{WangXS200804}. The electromagnetic aether behaves as a visco-elastic continuum \cite{WangXS200804}. Maxwell's equations approximate the macroscopic behavior of the $\Omega(1)$ particles, in analogy to the way that classical elastic mechanics approximates the macroscopic behavior of the atoms of solid materials.

Descartes interpreted the celestial motions of celestial bodies based on the hypothesis that the universe is filled by a fluidic vortex aether \cite{WhittakerE1951}. Since Newton's law of gravitation was published in 1687 \cite{NewtonI1962}, this action-at-a-distance theory was criticized by the French Cartesian \cite{WhittakerE1953}. Sir I. Newton tried to obtain a derivation of his law based on Descartes' scientific research program. At last, he proved that Descartes' vortex aether hypothesis could not explain celestial motions properly \cite{NewtonI1962}. Newton himself even suggested an explanation of gravity based on the action of an aetherial medium pervading the space (\cite{WhittakerE1951}, p.\ 28). Euler attempted to explain gravity based on some hypotheses of a fluidic aether \cite{WhittakerE1953}. Following Descartes, we introduce the following assumption \cite{WangXS200810}.
\begin{assumption_my}\label{assumption 300-150}
We suppose that vacuum is filled by an extremely thin medium which may be called the $\Omega(0)$ substratum, or the gravitational aether.
\end{assumption_my}

Thus, two sinks in the $\Omega(0)$ substratum are found to attract with each other according to the inverse-square law of gravitation \cite{WangXS200810}. A feature of this theory is that the gravitational constant depends on time and the location in space.

The particles that constitute the $\Omega(1)$ substratum may be called the $\Omega(1)$ particles.  Lord Kelvin believes that the electromagnetic aether must also generate gravity \cite{KelvinL1901,DupreMJ2012}. Following Lord Kelvin, we introduce the following assumption.
\begin{assumption_my}\label{assumption 300-200}
We suppose that there exist a kind of basic sinks of the $\Omega(0)$ substratum, which may be called monads after Leibniz. The $\Omega(1)$ particles and elementary particles are formed of monads.
\end{assumption_my}

Since monads, the $\Omega(1)$ particles, elementary particles are sinks of the $\Omega(0)$ substratum, they attract with each other according to Newton's law of gravitation \cite{WangXS200810}.

There exists an universal drag force exerted on each sink of the $\Omega(0)$ substratum \cite{WangXS200810}. Therefore, each monad, each $\Omega(1)$ particle and each elementary particle, as sinks of the $\Omega(0)$ substratum, will experience the universal drag force. On the other hand, all the monads, $\Omega(1)$ particles and elementary particles are undertaking stochastic movements. Based on this universal damping force and some assumptions, microscopic particles are found to obey a generalized non-relativistic Schr\"{o}dinger equation \cite{WangXS2014PhysicsEssays}.

For convenience, we may call these theories \cite{WangXS200804,WangXS200810,WangXS2014PhysicsEssays,WangXS201908} as the theory of vacuum mechanics. Vacuum mechanics is a physical theory which attempts to derive some basic physical laws based on a new mechanical model of vacuum and particles.

\section{Equivalence between the inertial mass and the gravitational mass \label{sec 820}}
\begin{proposition_my}\label{equivalence 820-50}
The inertial mass of a microscopic particle equals its gravitational mass.
\end{proposition_my}
{\bfseries{Proof of Proposition \ref{equivalence 820-50}.}} Newton's law of gravitation can be written as (\cite{OhanianHC2013}, p.\ 2)
\begin{equation}\label{Newton 820-100}
\mathbf{F}_{12}=-G_{N}\frac{m_{g1}m_{g2}}{r^2}\hat{\mathbf{r}}_{12},
\end{equation}
where $\mathbf{F}_{12}(t)$ denotes the force exerted on the particle with gravitational mass $m_{g2}$ by the particle with gravitational mass $m_{g1}$, $m_{g1}$ and $m_{g2}$ are the gravitational masses of two particles, $G_{N}$ is Newton's gravitational constant, $\hat{\mathbf{r}}_{12}$ denotes the unit vector directed along the line from the particle with mass $m_{g1}$ to the particle with mass $m_{g2}$, $r$ is the distance between the two particles.

In 2008, we show that the force $\mathbf{F}_{12}(t)$  exerted on the particle with inertial mass $m_{i2}(t)$
 by the velocity field of the $\Omega(0)$ substratum induced by the particle with inertial mass $m_{i1}(t)$
 is \cite{WangXS200810}
\begin{equation}\label{gravitation 820-200}
\mathbf{F}_{12}(t)=- \gamma_{N}(t)\frac{m_{i1}(t)
m_{i2}(t)}{r^2}\hat{\mathbf{r}}_{12},
\end{equation}
where
\begin{equation}\label{constant 820-210}
\gamma_{N}(t)=\frac{\rho_{0} q^2_0}{4\pi m^2_0(t)},
\end{equation}
$\rho_{0}$ is the density of the $\Omega(0)$ substratum, $m_0(t)$ is the inertial mass of a monad at time $t$, $-q_0( q_0 > 0)$ is the strength of the monad.

Suppose that $G_{N}=\gamma_{N}(t)$. Comparing Eq.\ (\ref{Newton 820-100}) and Eq.\ (\ref{gravitation 820-200}), we have
\begin{equation}\label{mass 820-200}
m_{i1}m_{i2}=m_{g1}m_{g2}.
\end{equation}

Now we study a gravitational system of two protons. According to Eq.\ (\ref{mass 820-200}), we have
\begin{equation}\label{mass 820-300}
m_{ip}^{2}=m_{gp}^{2},
\end{equation}
where $m_{ip}$ and $m_{gp}$ are the inertial mass and gravitational mass of a proton respectively.

Noticing $m_{ip}>0$ and $m_{gp}>0$, Eq.\ (\ref{mass 820-300}) can be written as
\begin{equation}\label{mass 820-400}
m_{ip}=m_{gp}.
\end{equation}

Eq.\ (\ref{mass 820-400}) shows that the inertial mass $m_{ip}$ of a proton equals its gravitational mass $m_{gp}$. Similarly, we can demonstrate that the inertial mass of another type of microscopic particle equals its gravitational mass. $\Box$

This result is called the principle of equivalence in the theory of general relativity \cite{MollerC1972,WeinbergS1972,MisnerC1973}.

\section{The dynamical gravitational potentials in inertial reference frames\label{sec 830}}
The purpose of this section is to review the mathematical forms of the dynamical gravitational potentials in inertial reference frames. These results may provide us some clues to explore possible mathematical models of inertial potential and inertial force Lagrangian in non-inertial reference frames, which are introduced in the next section.

We introduce a Cartesian coordinate system $\{ 0, x, y, z \}$ for a three-dimensional Euclidean space that attached to the $\Omega(1)$ substratum. Let $\{ 0, t \}$ be a one-dimensional time coordinate. For convenience, we introduce the following Galilean coordinate system
\begin{math}
x^{0} \equiv ct, \quad  \ x^{1} \equiv x, \quad   \  x^{2} \equiv y,  \quad  \  x^{3} \equiv z.
\end{math}
Let $\eta_{\mu \nu}$ denotes the metric tensor of the Minkowski spacetime.

We will use Greek indices $\alpha, \beta, \mu, \nu$, etc., to denote the range $\{0, 1, 2, 3\}$
and Latin indices $i, j, k$, etc., to denote the range $\{1, 2, 3\}$.
Einstein's summation convention will be used, i.e., any repeated Greek superscript or subscript appearing in a term of an equation is to be summed from $0$ to $3$.

The definition of the strength $\mathbf{g}$ of a gravitational field is (\cite{LiuL2004}, p.\ 24)
\begin{equation}\label{strength 820-500}
\mathbf{g}=\frac{\mathbf{F}_{g}}{m_{\mathrm{test}}},
\end{equation}
where $m_{\mathrm{test}}$ is the mass of a test point particle, $\mathbf{F}_{g}$ is the gravitational force exerted on
the test point particle by a gravitational field.

According to Newton's second law, we have
\begin{equation}\label{Newton 820-600}
\mathbf{F}_{g}=m_{\mathrm{test}}\mathbf{a},
\end{equation}
where $\mathbf{a}$ is the acceleration of the test point particle.

Comparing Eq.\ (\ref{Newton 820-600}) and Eq.\ (\ref{strength 820-500}), we have
\begin{equation}\label{strength 820-700}
\mathbf{g}=\mathbf{a}.
\end{equation}

The definition of the acceleration $\mathbf{a}$ is (\cite{LiuL2004}, p.\ 24)
\begin{equation}\label{acceleration 820-800}
{a}_{i}=\gamma_{ik}\frac{d^{2}x^{k}}{dt^2},
\end{equation}
where
\begin{equation}\label{metric 820-900}
\gamma_{ik}=-g_{ik}+\frac{g_{0i}g_{0k}}{g_{00}},
\end{equation}
$g_{\mu \nu}$ is the metric tensor of a Riemannian spacetime.

Based on the time tracks of free particles described by geodesic curves in Minkowski spacetime, we have the following results
(\cite{MollerC1972}, p.\ 279;\cite{LiuL2004}, p.\ 26)
\begin{equation}\label{acceleration 820-1000}
{a}_{i}=-\frac{\partial \Pi}{\partial x^{i}}-c\sqrt{1+\frac{2\Pi}{c^{2}}}\frac{\partial \gamma_{i}}{\partial t},
\end{equation}
where
\begin{equation}\label{potential 820-1100}
\Pi=-\frac{1-g_{00}}{2}c^{2}, \quad \gamma_{i}=-\frac{g_{i0}}{\sqrt{g_{00}}}.
\end{equation}

If we suppose that $\partial g_{00} / \partial t \approx 0$, then Eqs.\ (\ref{acceleration 820-1000}) can also be written as
\begin{equation}\label{acceleration 820-1200}
{a}_{i}=c^{2}\frac{\partial }{\partial x^{i}}\left ( \frac{1-g_{00}}{2}\right )-c^{2}\frac{\partial (-g_{i0})}{\partial (ct)}.
\end{equation}

\section{Inertial potential and inertial force Lagrangian in non-inertial reference frames\label{sec 850}}
According to the theory of general relativity \cite{WeinbergS1972,MisnerC1973}, the Einstein's equations are valid not only in inertial reference frames but also in non-inertial reference frames. Thus, it is needed to explore the possibility to derive the Einstein's equations in non-inertial reference frames.

When solving the Einstein's equations for an isolated system of masses, V. Fock introduces harmonic reference frame and obtains an unambiguous solution (\cite{FockV1964}, p.\ 369).  Furthermore, in the case of an isolated system of masses, he concludes that there exists a harmonic reference frame which is determined uniquely apart from a Lorentz transformation if suitable supplementary conditions are imposed (\cite{FockV1964}, p.\ 373). It is known that wave equations keep the same form under Lorentz transformations \cite{MollerC1972}. From Eqs.\ (93) in Ref. \cite{WangXS201908}, we notice that the field equations of gravitation in inertial reference frames in a Minkowski spacetime are wave equations. Thus, we speculate that Fock's special harmonic reference frames may have provided us a clue to derive the Einstein's equations in some special class of non-inertial reference frames.

We introduce an arbitrary non-inertial coordinate system $(x'^{0}, \ x'^{1},  \  x'^{2},  \  x'^{3})$ and denote it as $S_{n}$. It is known that a particle in a non-inertial reference frame will experience an inertial force. Unfortunately, we have no knowledge about the origin of inertial forces.

For convenience, we introduce the following definition of matter system.
\begin{definition_my}\label{matter 850-50}
A matter system is a system of a number of elementary particles, or continuously distributed elementary particles.
\end{definition_my}

The equivalence between inertial mass and gravitational mass implies that to some degree gravitational forces behave in the same way as inertial forces (\cite{OhanianHC2013}, p.\ 17). Thus, we speculate that inertial forces may originate from the interactions between matter systems and vacuum.  Therefore, we introduce the following assumption.
\begin{assumption_my}\label{inertial 850-50}
The inertial force exerted on a matter system in a non-inertial reference frame stems from the interactions between the matter system and vacuum.
\end{assumption_my}

Based on Assumption \ref{inertial 850-50}, we introduce the following concepts for inertial forces, which are similar to those concepts for gravitational interactions.
\begin{definition_my}\label{potential 850-100}
Inertial potential $\psi^{\mathrm{iner}}_{\mu\nu}$ is an interaction potential between a matter system and vacuum resulting from the inertial force $\mathbf{F}_{\mathrm{iner}}$ exerted on the matter system by vacuum in a non-inertial reference frame $S_{n}$.
\end{definition_my}

\begin{definition_my}\label{Lagrangian 850-200}
Inertial force Lagrangian $L_{\mathrm{iner}}$ is an interaction Lagrangian between a matter system and vacuum resulting from the inertial force $\mathbf{F}_{\mathrm{iner}}$ exerted on the matter system by vacuum in a non-inertial reference frame $S_{n}$.
\end{definition_my}

Now our task is to explore possible models of inertial potential $\psi^{\mathrm{iner}}_{\mu\nu}$ and inertial force Lagrangian $L_{\mathrm{iner}}$. Let $\eta'_{\mu\nu}$ denotes the metric tensor of the non-inertial reference frame $S_{n}$. Suppose that $\partial \eta'_{00} / \partial x'^{0} \approx 0$. Then, following similar methods as in the derivation of Eqs.\ (\ref{acceleration 820-1200}), we obtain the following relationship for the inertial acceleration $\mathbf{a}$ of a test point particle in the non-inertial reference frame $S_{n}$
\begin{equation}\label{acceleration 850-300}
{a}_{i}=c^{2}\frac{\partial }{\partial x'^{i}}\left ( \frac{1-\eta'_{00}}{2}\right )-c^{2}\frac{\partial (-\eta'_{i0})}{\partial x'^{0}}.
\end{equation}

If $\eta'_{i0}$ are time-independent, the inertial acceleration $\mathbf{a}$ of the test point particle in Eqs.\ (\ref{acceleration 850-300}) simplifies to (\cite{MollerC1972}, p.\ 280)
\begin{equation}\label{acceleration 850-400}
{a}_{i}=c^{2}\frac{\partial }{\partial x'^{i}}\left ( \frac{1-\eta'_{00}}{2}\right ).
\end{equation}

Using Eqs.\ (\ref{acceleration 850-400}), the inertial force $\mathbf{F}_{\mathrm{iner}}$ exerted on the test point particle can be written as
\begin{equation}\label{force 850-500}
\mathbf{F}_{\mathrm{iner}} = m\mathbf{a} = mc^{2}\nabla \left ( \frac{1-\eta'_{00}}{2}\right ),
\end{equation}
where $m$ is the mass of the test point particle, $\nabla =
\mathbf{i}\partial /\partial x'^{1} +
\mathbf{j}\partial /\partial x'^{2} +
\mathbf{k}\partial /\partial x'^{3}$ is the
Hamilton operator, $\mathbf{i}$, $\mathbf{j}$, $\mathbf{k}$ are three unit vectors directed along the
coordinate axes.

From Eq.\ (\ref{force 850-500}), the inertial force Lagrangian of a system of vacuum and the test point particle can be written as
\begin{equation}\label{Lagrangian 850-600}
L_{\mathrm{iner1}}=-mc^{2} \left ( \frac{1-\eta'_{00}}{2}\right ).
\end{equation}

Therefore, the inertial force Lagrangian of a system of vacuum and continuously distributed particles can be written as
\begin{equation}\label{Lagrangian 850-700}
L_{\mathrm{iner}}=-\rho_{m}c^{2} \left ( \frac{1-\eta'_{00}}{2}\right ),
\end{equation}
where $\rho_{m}$ is the rest mass density of the continuously distributed particles.

Let $T_{m}^{'\mu\nu}$ denotes the contravariant energy-momentum tensor of the continuously distributed particles system. If the inertial force is small enough, then we may regard this non-inertial reference frame $S_{n}$ as an inertial reference frame approximately. Thus, we have $T_{m}^{'00} \approx \rho_{m}c^{2}$ \cite{WangXS201908}. Noticing $\eta_{00} = 1$, the inertial force Lagrangian $L_{\mathrm{iner}}$ in Eq.\ (\ref{Lagrangian 850-700}) can be written as
\begin{equation}\label{Lagrangian 850-800}
L_{\mathrm{iner}}\approx f_{0}\psi^{\mathrm{iner}}_{00}T_{m}^{'00},
\end{equation}
where
\begin{equation}\label{potential 850-900}
\psi^{\mathrm{iner}}_{00}=-\frac{1}{2f_{0}}(\eta_{00}-\eta'_{00}).
\end{equation}

Following Ref. \cite{WangXS201908}, the parameter $f_{0}$ is
\begin{equation}\label{f0 850-950}
f_{0}=\sqrt{\frac{2\rho_{0}q_{0}^{2}}{m_{0}^{2}c^{4}}}=\sqrt{\frac{8\pi \gamma_{N}}{c^{4}}}.
\end{equation}

Inspired by Eq.\ (\ref{Lagrangian 850-800}) and Eq.\ (\ref{potential 850-900}), we introduce the following assumption.
\begin{assumption_my}\label{Lagrangian 870-150}
Suppose that the inertial force Lagrangian $L_{\mathrm{iner}}$ of a system of a free point particle and vacuum in the non-inertial reference frame $S_{n}$ can be written as
\begin{equation}\label{Lagrangian 870-210}
L_{\mathrm{iner}}=f_{0}\psi^{\mathrm{iner}}_{\mu\nu} m u'^{\mu}u'^{\nu},
\end{equation}
where $m$ is the rest mass of the point particle, $u'^{\mu}\equiv dx^{'\mu}/d\tau_{\eta'}$, $\tau_{\eta'}$ is the proper time,
\begin{equation}\label{tensor 870-220}
\psi^{\mathrm{iner}}_{\mu\nu}=-\frac{1}{2f_{0}}(\eta_{\mu\nu}-\eta'_{\mu\nu}).
\end{equation}
\end{assumption_my}

Following similar methods as in \cite{WangXS201908}, we obtain the following result.
\begin{proposition_my}\label{motion 870-200}
Suppose that Assumption \ref{Lagrangian 870-150} is valid. Then, the equations of motion of a free point particle can be written as
\begin{equation}\label{motion 870-230}
\frac{d^{2}x^{'\mu}}{d\tau_{\eta'}^{2}}+C_{\nu\sigma}^{'\mu}\frac{dx^{'\nu}}{d \tau_{\eta'}}\frac{dx^{'\sigma}}{d\tau_{\eta'}}= 0,
\end{equation}
where
\begin{equation}\label{Christoffel 870-235}
C_{\alpha\beta}^{'\nu}\equiv\frac{1}{2}\eta^{'\mu\nu}
\left ( \frac{\partial \eta'_{\mu\alpha}}{\partial x^{'\beta}}
+\frac{\partial \eta'_{\mu\beta}}{\partial x^{'\alpha}}
-\frac{\partial \eta'_{\alpha\beta}}{\partial x^{'\mu}}\right )
\end{equation}
are the corresponding Christoffel symbols in the non-inertial reference frame $S_{n}$.
\end{proposition_my}
{\bfseries{Proof of Proposition \ref{motion 870-200}.}} The Lagrangian of a free point particle in $S_{n}$ can be written as (\cite{FeynmanRP1995}, p.\ 57;\cite{ThirringW1961})
\begin{equation}\label{Lagrangian 870-240}
L'_{0} = \frac{1}{2}m\frac{dx^{'\mu}}{d\tau_{\eta'}}\frac{dx'_{\mu}}{d\tau_{\eta'}}=\frac{1}{2}mu^{'\mu}u'_{\mu}=\frac{1}{2}m\eta'_{\mu\nu}u^{'\mu}u^{'\nu},
\end{equation}
where $m$ is the rest mass of the point particle, $d\tau_{\eta'} \equiv \frac{1}{c}\sqrt{dx^{'\mu}dx'_{\mu}}$ is the infinitesimal proper time interval, $u^{'\mu}\equiv dx^{'\mu}/d\tau_{\eta'}$.

Ignoring the contravariant energy-momentum tensor of the $\Omega(1)$ substratum, i.e.,  $T_{\Omega(1)}^{'\mu\nu}\approx 0$ and using Eq.\ (\ref{Lagrangian 870-240}) and Eq.\ (\ref{Lagrangian 870-210}), the total Lagrangian $L_{\mathrm{p}}$ of a system of the $\Omega(0)$ substratum, the $\Omega(1)$ substratum and the point particle can be written as
\begin{equation}\label{Lagrangian 870-250}
L_{\mathrm{p}}=L'_{0}+L_{\mathrm{iner}}=\frac{1}{2}mu'^{\mu}u'_{\mu}+f_{0}\psi^{\mathrm{iner}}_{\mu\nu} m u'^{\mu}u'^{\nu}.
\end{equation}

The Euler-Lagrange equations for the total Lagrangian $L_{\mathrm{p}}$ can be written as (\cite{OhanianHC2013}, p.\ 111)
\begin{equation}\label{Euler 870-260}
\frac{\partial L_{\mathrm{p}}}{\partial x^{'\mu}}-\frac{d}{d\tau_{\eta'}}\frac{\partial L_{\mathrm{p}}}{\partial u'^{\mu}}=0.
\end{equation}

Putting Eq.\ (\ref{Lagrangian 870-250}) into Eqs.\ (\ref{Euler 870-260}), we have
\begin{eqnarray}
&&\frac{d}{d\tau_{\eta'}}\left [ \left (\eta_{\mu\nu}+2f_{0}\psi^{\mathrm{iner}}_{\mu\nu}\right )\frac{dx^{'\nu}}{d\tau_{\eta'}}\right ] \nonumber\\
&&- f_{0}\frac{\partial \psi^{\mathrm{iner}}_{\alpha\beta}}{\partial x^{'\mu}}\frac{dx^{'\alpha}}{d \tau_{\eta'}}\frac{dx^{'\beta}}{d\tau_{\eta'}} = 0. \label{motion 870-270}
\end{eqnarray}

Using Eq.\ (\ref{tensor 870-220}), Eqs.\ (\ref{motion 870-270}) can be written as
\begin{equation}\label{motion 870-280}
\frac{d}{d\tau_{\eta'}}\left (\eta'_{\mu\nu}\frac{dx^{'\nu}}{d\tau_{\eta'}}\right ) - \frac{1}{2}\frac{\partial \eta'_{\alpha\beta}}{\partial x^{'\mu}}\frac{dx^{'\alpha}}{d \tau_{\eta'}}\frac{dx^{'\beta}}{d\tau_{\eta'}} = 0.
\end{equation}

Eqs.\ (\ref{motion 870-280}) represent a geodesic line in a Minkowski spacetime with a metric tensor $\eta'_{\mu\nu}$, which can also be written as Eqs.\ (\ref{motion 870-230}) (\cite{LiuL2004}, p.\ 50). $\Box$

Eqs.\ (\ref{motion 870-230}) is a geodesic curve in a Minkowski spacetime. It is known that a geodesic curve is a straight line in a Minkowski spacetime (\cite{FaraoniV2013}, p.\ 235). For instance, according to Newton's first law, a free particle moves along a straight line in the Galilean coordinates. Therefore, Assumption \ref{Lagrangian 870-150} may be supported by some experiments. Inspired by the inertial force Lagrangian for a free point particle in Eq.\ (\ref{Lagrangian 870-210}), we introduce the following assumption for a matter system and vacuum based on Assumption \ref{assumption 300-200}.

\begin{assumption_my}\label{assumption 870-300}
The inertial force Lagrangian $L_{\mathrm{iner}}$ of a matter system and vacuum in the non-inertial reference frame $S_{n}$ can be written as
\begin{equation}\label{inertial 870-310}
L_{\mathrm{iner}}=f_{0}\psi^{\mathrm{iner}}_{\mu\nu}(T_{m}^{'\mu\nu}+T_{\Omega(1)}^{'\mu\nu})+O[(f_{0}\psi^{\mathrm{iner}}_{\mu\nu})^{2}],
\end{equation}
where $T_{m}^{'\mu\nu}$ and $T_{\Omega(1)}^{'\mu\nu}$ are the contravariant energy-momentum tensors of the matter and the $\Omega(1)$ substratum respectively, $O[(f_{0}\psi^{\mathrm{iner}}_{\mu\nu})^{2}]$ denotes those terms which are small quantities of the order of $(f_{0}\psi^{\mathrm{iner}}_{\mu\nu})^{2}$.
\end{assumption_my}

\section{Field equations in a special class of non-inertial reference frames \label{sec 870}}
Suppose that the transformation equations between a non-inertial coordinate system $(x'^{0}, \ x'^{1},  \  x'^{2},  \  x'^{3})$ and the Galilean coordinates $(ct, x, y, z)$ are
\begin{equation}\label{transformation 870-50}
x'^{\alpha}= f^{\alpha}(x^{0}, \ x^{1},  \  x^{2},  \  x^{3}).
\end{equation}

Following V. Fock (\cite{FockV1964}, p.\ 370-373), we introduce the following definition of a special class of reference frames.
\begin{definition_my}\label{Fock 870-100}
Suppose that a coordinate system $(x'^{0}, \ x'^{1},  \  x'^{2},  \  x'^{3})$ satisfies the following conditions:
(1) every coordinates $x'^{\alpha}$ satisfies the d'Alembert's equation (\cite{FockV1964}, p.\ 369), i.e.,
\begin{equation}\label{harmonic 870-110}
\square_{\eta'}x'^{\alpha} \equiv \frac{1}{\sqrt{-\eta'_{0}}}\frac{\partial }{\partial x'^{\mu}}
\left ( \sqrt{-\eta'_{0}}\eta'^{\mu\nu}\frac{\partial x'^{\alpha}}{\partial x'^{\nu}}\right )= 0,
\end{equation}
where $\eta'_{\mu\nu}$ is the metric of the reference frame $S_{n}$, $\eta'_{0} = \mathrm{Det} \ \eta'_{\mu\nu}$, $\mathrm{Det} A_{\mu\nu}$ denotes the value of the corresponding determinant of the tensor $A_{\mu\nu}$;
(2) every coordinates $x'^{\alpha}$ converges to the Galilean coordinates $(ct, x, y, z)$ at large enough distance, i.e.,
\begin{equation}\label{Galilean 870-120}
\lim_{r \rightarrow \infty}x'^{\alpha}= x^{\alpha},
\end{equation}
where $r=\sqrt{x^{2}+y^{2}+z^{2}}$;
(3) $\eta'^{\mu\nu}-(\eta'^{\mu\nu})_{\infty}$ are outgoing waves, i.e., $\eta'^{\mu\nu}-(\eta'^{\mu\nu})_{\infty}$ satisfy the following condition of outward radiation: for $r \rightarrow \infty$, and all values of $t'_{0}=t+r/c$ in an arbitrary fixed interval the following limiting conditions are satisfied (\cite{FockV1964}, p.\ 365)
\begin{eqnarray}
&&\lim_{r \rightarrow \infty}\left [ \frac{\partial [r(\eta'^{\mu\nu}-(\eta'^{\mu\nu})_{\infty})]}{\partial r} \right.\nonumber\\
&&\left.+ \frac{1}{c}\frac{\partial [r(\eta'^{\mu\nu}-(\eta'^{\mu\nu})_{\infty})]}{\partial t}\right ] = 0, \label{outgoing 870-130}
\end{eqnarray}

where $(\eta'^{\mu\nu})_{\infty}$ denotes the value of $\eta'^{\mu\nu}$ at infinity.
Then, we call this coordinate system $(x'^{0}, \ x'^{1},  \  x'^{2},  \  x'^{3})$ as a Fock coordinate system.
\end{definition_my}

We use $S_{F}$ to denote a Fock coordinate system. The Galilean coordinate system $(ct, x, y, z)$ is a Fock coordinate system. V. Fock points out an advantage of Fock coordinate system (\cite{FockV1964}, p.\ 369):"{\itshape  When solving Einstein's equations for an isolated system of masses we used harmonic coordinates and in this way obtained a perfectly unambiguous solution.}"  Here the harmonic coordinates called by V. Fock are Fock coordinate systems.

According to a theorem of Fock about Fock coordinate systems (\cite{FockV1964}, p.\ 369-373), the transformation equations (\ref{transformation 870-50}) can be written as a Lorentz transformation, i.e.,
\begin{equation}\label{transformation 870-150}
x'^{\mu}=a^{\mu}_{\  \nu}x^{\nu},
\end{equation}
where $a^{\mu}_{\  \nu}$ are coefficients of a Lorentz transformation.

For convenience, we introduce the following notations
\begin{equation}\label{notation 870-350}
\partial'_{\mu}\equiv\left (\frac{\partial}{\partial x^{'0}}, \frac{\partial}{\partial x^{'1}}, \frac{\partial}{\partial x^{'2}}, \frac{\partial}{\partial x^{'3}}\right ), \quad  \partial^{'\mu} \equiv \eta^{'\mu\nu}\partial'_{\nu}.
\end{equation}

\begin{proposition_my}\label{Lagrangian 870-400}
Suppose that the reference frame $S_{F}$ is a Fock coordinate system and Assumptions \ref{assumption 870-300} is valid, then the total Lagrangian $L'_{\mathrm{tot}}$ of a system of the $\Omega(0)$ substratum, the $\Omega(1)$ substratum, vacuum and matter in $S_{F}$ can be written as
\begin{eqnarray}
L'_{\mathrm{tot}} &= &\frac{1}{2}\partial'_{\lambda}\psi'_{\mu\nu}\partial'^{\lambda}\psi^{'\mu\nu} -2\partial'_{\lambda}\psi'_{\mu\nu}\partial^{'\mu}\psi^{'\lambda\nu} \nonumber\\
&&-6\partial^{'\mu}\psi'_{\mu\nu}\partial^{'\nu}\psi'-\frac{3}{2} \partial'_{\lambda}\psi'\partial^{'\lambda}\psi' \nonumber\\
&&+L'_{\mathrm{more}} +f_{0}\psi'_{\mu\nu}(T_{m}^{'\mu\nu}+T_{\Omega(1)}^{'\mu\nu})\nonumber\\
&&+f_{0}\psi^{\mathrm{iner}}_{\mu\nu}(T_{m}^{'\mu\nu}+T_{\Omega(1)}^{'\mu\nu})\nonumber\\
&&+O[(f_{0}\psi^{\mathrm{iner}}_{\mu\nu})^{2}]+O[(f_{0}\psi'_{\mu\nu})^{2}],\label{Lagrangian 870-410}
\end{eqnarray}
where $\psi'_{\mu\nu}$ is a tensorial potential of the gravitational field, $L'_{\mathrm{more}}$ denotes those terms involving more than two derivatives of $\psi'_{\mu\nu}$, $O[(f_{0}\psi'_{\mu\nu})^{2}]$ denotes those terms which are small quantities of the order of $(f_{0}\psi'_{\mu\nu})^{2}$.
\end{proposition_my}
{\bfseries{Proof of Proposition \ref{Lagrangian 870-400}.}}
Based on some assumptions, the total Lagrangian $L_{\mathrm{tot}}$ of a system of the $\Omega(0)$ substratum, the $\Omega(1)$ substratum and matter in an inertial reference frame can be written as \cite{WangXS201908}
\begin{eqnarray}
L_{\mathrm{tot}} &= &\frac{1}{2}\partial_{\lambda}\psi_{\mu\nu}\partial^{\lambda}\psi^{\mu\nu} -2\partial_{\lambda}\psi_{\mu\nu}\partial^{\mu}\psi^{\lambda\nu} \nonumber\\
&&-6\partial^{\mu}\psi_{\mu\nu}\partial^{\nu}\psi-\frac{3}{2} \partial_{\lambda}\psi\partial^{\lambda}\psi +L_{\mathrm{more}} \nonumber\\
&&+f_{0}\psi_{\mu\nu}(T_{m}^{\mu\nu}+T_{\Omega(1)}^{\mu\nu})\nonumber\\
&&+O[(f_{0}\psi_{\mu\nu})^{2}],\label{Lagrangian 870-3300}
\end{eqnarray}
where $\psi_{\mu\nu}$ is a tensorial potential of the gravitational field in the inertial reference frame, $L_{\mathrm{more}}$ denotes those terms involving more than two derivatives of $\psi_{\mu\nu}$, $O[(f_{0}\psi_{\mu\nu})^{2}]$ denotes those terms which are small quantities of the order of $(f_{0}\psi_{\mu\nu})^{2}$.

The total Lagrangian $L'_{\mathrm{tot}}$ can be written as
\begin{equation}\label{Lagrangian 870-450}
L'_{\mathrm{tot}}=L_{\mathrm{tot}}+L_{\mathrm{iner}}.
\end{equation}

Similar to the case of inertial reference frames (\cite{OhanianHC2013}, p.\ 59-60, 63), we also have the following results in the Fock coordinate system $S_{F}$
\begin{eqnarray}
&&\partial'_{\lambda}=a^{\ \sigma}_{\lambda}\partial_{\sigma}, \quad   \partial^{'\lambda}=a^{\lambda}_{\ \sigma}\partial^{\sigma}, \label{transform 870-500}\\
&&\psi^{'\mu\nu}=a^{\mu}_{\ \alpha}a^{\nu}_{\ \beta}\psi^{\alpha\beta},  \label{transform 870-610}\\
&&\psi'_{\mu\nu}=a^{\ \alpha}_{\mu}a^{\ \beta}_{\nu}\psi_{\alpha\beta}.  \label{transform 870-620}
\end{eqnarray}

The first term on the right hand side of Eqs.\ (\ref{Lagrangian 870-410}) can be written as
\begin{eqnarray}
\frac{1}{2}\partial'_{\lambda}\psi'_{\mu\nu}\partial'^{\lambda}\psi^{'\mu\nu} &=& \frac{1}{2}(a^{\ \sigma}_{\lambda}\partial_{\sigma}) (a^{\ \alpha}_{\mu}a^{\ \beta}_{\nu}\psi_{\alpha\beta}) \nonumber\\
&&\cdot(a^{\lambda}_{\ \sigma}\partial^{\sigma})(a^{\mu}_{\ \alpha}a^{\nu}_{\ \beta}\psi^{\alpha\beta}).\ \ \ \ \label{term 870-700}
\end{eqnarray}

We have the following result (\cite{OhanianHC2013}, p.\ 60)
\begin{equation}\label{Lorentz 870-800}
a^{\ \mu}_{\beta} a^{\beta}_{\ \nu}=\delta_{\nu}^{\mu},
\end{equation}
where $\delta_{\nu}^{\mu}$ is the Kronecker delta.

Using Eq.\ (\ref{Lorentz 870-800}), Eqs.\ (\ref{term 870-700}) can be written as
\begin{equation}\label{transform 870-900}
\frac{1}{2}\partial'_{\lambda}\psi'_{\mu\nu}\partial'^{\lambda}\psi^{'\mu\nu}
=\frac{1}{2}\partial_{\sigma}\psi_{\alpha\beta}\partial^{\sigma}\psi^{\alpha\beta}.
\end{equation}

Similarly, we can verify the following results
\begin{eqnarray}
-2\partial'_{\lambda}\psi'_{\mu\nu}\partial^{'\mu}\psi^{'\lambda\nu}
&=&-2\partial_{\sigma}\psi_{\alpha\beta}\partial^{\alpha}\psi^{\sigma\beta}, \label{transform 870-910}\\
-6\partial^{'\mu}\psi'_{\mu\nu}\partial^{'\nu}\psi'&=&-6\partial^{\alpha}\psi_{\alpha\beta}\partial^{\beta}\psi, \label{transform 870-920}\\
-\frac{3}{2} \partial'_{\lambda}\psi'\partial^{'\lambda}\psi'&=&-\frac{3}{2} \partial_{\sigma}\psi\partial^{\sigma}\psi, \label{transform 870-930}\\
f_{0}\psi'_{\mu\nu}T_{m}^{'\mu\nu}&=&f_{0}\psi_{\alpha\beta}T_{m}^{\alpha\beta}, \label{transform 870-940}\\
L'_{\mathrm{more}}&=&L_{\mathrm{more}}, \label{transform 870-950}\\
O[(f_{0}\psi'_{\mu\nu})^{2}]&=&O[(f_{0}\psi_{\mu\nu})^{2}]. \label{transform 870-960}
\end{eqnarray}

Putting Eq.\ (\ref{Lagrangian 870-3300}) and Eq.\ (\ref{inertial 870-310}) into Eq.\ (\ref{Lagrangian 870-450}) and using Eqs.\ (\ref{transform 870-900}-\ref{transform 870-960}), we obtain Eq.\ (\ref{Lagrangian 870-410}). $\Box$

Applying similar methods as in Ref. \cite{WangXS201908}, we have the following result.
\begin{theorem_my}\label{field 870-1000}
If we ignore those terms which are small quantities of the order of $(f_{0}\psi'_{\mu\nu})^{2}$ and $(f_{0}\psi^{\mathrm{iner}}_{\mu\nu})^{2}$ and those terms involving more than two derivatives of $\psi'_{\mu\nu}$ in Eq.\ (\ref{Lagrangian 870-410}), i.e., $O[(f_{0}\psi'_{\mu\nu})^{2}] \approx 0$, $O[(f_{0}\psi^{\mathrm{iner}}_{\mu\nu})^{2}] \approx 0$ and $L'_{\mathrm{more}} \approx 0$, then the field equations for the total Lagrangian $L'_{\mathrm{tot}}$ in Eq.\ (\ref{Lagrangian 870-410}) can be written as
\begin{eqnarray}
&&\partial'_{\sigma}\partial'^{\sigma}\psi'_{\alpha\beta}-2(\partial^{'\sigma}\partial'_{\alpha}\psi'_{\beta\sigma}
+\partial^{'\sigma}\partial'_{\beta}\psi'_{\alpha\sigma})\nonumber\\
&&-6(\eta'_{\alpha\beta}\partial'_{\sigma}\partial'_{\lambda}\psi^{'\sigma\lambda}
+\partial'_{\alpha}\partial'_{\beta}\psi')-3\eta'_{\alpha\beta}\partial'_{\sigma}\partial^{'\sigma}\psi'\nonumber\\
&&=f_{0}\left (T^{'m}_{\alpha\beta}+T^{'\Omega(1)}_{\alpha\beta}\right ).\label{field 870-1010}
\end{eqnarray}
\end{theorem_my}
{\bfseries{Proof of Theorem \ref{field 870-1000}.}} We have the following Euler-Lagrange equations \cite{BabakSV1999}
\begin{equation}\label{field 1300-100}
\frac{\partial L'_{\mathrm{tot}}}{\partial \psi^{'\alpha\beta}}-\frac{\partial }{\partial x^{'\sigma}}\left ( \frac{\partial L'_{\mathrm{tot}}}{\partial (\partial'_{\sigma}\psi^{'\alpha\beta})}\right )=0.
\end{equation}

We have the following results
\begin{eqnarray}
\frac{\partial (\partial'_{\lambda}\psi'_{\mu\nu}\partial^{'\lambda}\psi^{'\mu\nu})}{\partial (\partial'_{\sigma}\psi^{'\alpha\beta})}&=& \frac{\partial (\partial'_{\lambda}\psi'_{\mu\nu})}{\partial (\partial'_{\sigma}\psi^{'\alpha\beta})}\partial^{'\lambda}\psi^{'\mu\nu}\nonumber\\
&+&\partial'_{\lambda}\psi'_{\mu\nu}\frac{\partial (\partial^{'\lambda}\psi^{'\mu\nu})}{\partial (\partial'_{\sigma}\psi^{'\alpha\beta})}, \ \ \ \ \label{partial 1300-200}
\end{eqnarray}

\begin{equation}\label{trans 1300-300}
\psi'_{\mu\nu}=\eta'_{\mu\rho}\eta'_{\nu\tau}\psi^{'\rho\tau}.
\end{equation}

Using Eqs.\ (\ref{trans 1300-300}), we have
\begin{eqnarray}
\frac{\partial (\partial'_{\lambda}\psi'_{\mu\nu})}{\partial (\partial'_{\sigma}\psi^{'\alpha\beta})}&=&\frac{\partial }{\partial (\partial'_{\sigma}\psi^{'\alpha\beta})}\left ( \eta'_{\mu\rho}\eta'_{\nu\tau}\partial'_{\lambda}\psi^{'\rho\tau}\right )\nonumber\\
&=&\eta'_{\mu\rho}\eta'_{\nu\tau} \frac{\partial (\partial'_{\lambda}\psi^{'\rho\tau})}{\partial (\partial'_{\sigma}\psi^{'\alpha\beta})}\nonumber\\
&=&\eta'_{\mu\rho}\eta'_{\nu\tau} \delta^{\lambda}_{\sigma}\delta^{\rho}_{\alpha}\delta^{\tau}_{\beta}\nonumber\\
&=&\eta'_{\mu\alpha}\eta'_{\nu\beta}\delta^{\lambda}_{\sigma}.\label{partial 1300-400}
\end{eqnarray}

Using Eqs.\ (\ref{partial 1300-400}) and Eqs.\ (\ref{trans 1300-300}), the first term on the right hand side of Eqs.\ (\ref{partial 1300-200}) can be written as
\begin{equation}\label{term 1300-500}
 \frac{\partial (\partial'_{\lambda}\psi'_{\mu\nu})}{\partial (\partial'_{\sigma}\psi^{'\alpha\beta})}\partial^{'\lambda}\psi^{'\mu\nu}
=\partial^{'\sigma}\psi'_{\alpha\beta}.
\end{equation}

Similarly, the second term on the right hand side of Eqs.\ (\ref{partial 1300-200}) can be written as
\begin{equation}\label{term 1300-600}
\partial'_{\lambda}\psi'_{\mu\nu}\frac{\partial (\partial^{'\lambda}\psi^{'\mu\nu})}{\partial (\partial'_{\sigma}\psi^{'\alpha\beta})}
=\partial^{'\sigma}\psi'_{\alpha\beta}.
\end{equation}

Using Eqs.\ (\ref{partial 1300-200}), Eqs.\ (\ref{term 1300-500}) and Eqs.\ (\ref{term 1300-600}), we have
\begin{equation}\label{term 1300-700}
\frac{\partial }{\partial x^{'\sigma}}\left [ \frac{\partial (\partial'_{\lambda}\psi'_{\mu\nu}\partial^{'\lambda}\psi^{'\mu\nu})}{\partial (\partial'_{\sigma}\psi^{'\alpha\beta})}\right ]=2\partial'_{\sigma}\partial^{'\sigma}\psi'_{\alpha\beta}.
\end{equation}

Similarly, we can verify the following results
\begin{eqnarray}
\frac{\partial }{\partial x^{'\sigma}}\left [ \frac{\partial (\partial'_{\lambda}\psi'_{\mu\nu}\partial^{'\mu}\psi^{'\lambda\nu})}{\partial (\partial'_{\sigma}\psi^{'\alpha\beta})}\right ]&=&\partial^{'\sigma}\partial'_{\alpha}\psi'_{\beta\sigma}\nonumber\\
&&+\partial^{'\sigma}\partial'_{\beta}\psi'_{\alpha\sigma}, \label{term 1300-800}\\
\frac{\partial }{\partial x^{'\sigma}}\left [ \frac{\partial (\partial^{'\mu}\psi'_{\mu\nu}\partial^{'\nu}\psi')}{\partial (\partial'_{\sigma}\psi^{'\alpha\beta})}\right ]&=&\partial'_{\alpha}\partial'_{\beta}\psi'\nonumber\\
&+&\eta'_{\alpha\beta}\partial'_{\sigma}\partial'_{\lambda}\psi^{'\sigma\lambda}, \label{term 1300-900}\\
\frac{\partial }{\partial x^{'\sigma}}\left [ \frac{\partial (\partial'_{\lambda}\psi'\partial^{'\lambda}\psi')}{\partial (\partial'_{\sigma}\psi^{'\alpha\beta})}\right ]&=&2\eta'_{\alpha\beta}\partial'_{\sigma}\partial^{'\sigma}\psi', \label{term 1300-1000}\\
\frac{\partial L'_{\mathrm{tot}}}{\partial \psi^{'\alpha\beta}}&=&f_{0}(T^{'m}_{\alpha\beta}+T^{'\Omega(1)}_{\alpha\beta}).\ \ \ \ \ \label{term 1300-1100}
\end{eqnarray}

Putting Eq.\ (\ref{Lagrangian 870-410}) into Eqs.\ (\ref{field 1300-100}) and using Eqs.\ (\ref{term 1300-700}-\ref{term 1300-1100}), we obtain Eqs.\ (\ref{field 870-1010}). $\Box$

Following Ref. \cite{WangXS201908}, we introduce the following notation in the Fock coordinate system $S_{F}$.
\begin{eqnarray}
&&\Psi^{'\mu\nu}=\partial'_{\lambda}\partial^{'\lambda}\psi^{'\mu\nu}-2\partial'_{\lambda}\partial^{'\mu}\psi^{'\nu\lambda}
-2\partial'_{\lambda}\partial^{'\nu}\psi^{'\mu\lambda}\nonumber\\
&&-6\eta^{'\mu\nu}\partial'_{\sigma}\partial'_{\lambda}\psi^{'\sigma\lambda}
-6\partial^{'\mu}\partial^{'\nu}\psi'-3\eta^{'\mu\nu}\partial'_{\lambda}\partial^{'\lambda}\psi'. \ \ \ \ \ \label{Psi 870-1100}
\end{eqnarray}

For convenience, we introduce the following definition of the total energy-momentum tensor $T^{'\mu\nu}$ of the system of the matter, the $\Omega(1)$ substratum and the $\Omega(0)$ substratum in the Fock coordinate system $S_{F}$
\begin{equation}\label{tensor 870-1200}
T^{'\mu\nu}=T_{m}^{'\mu\nu}+T_{\Omega(1)}^{'\mu\nu}+T_{\Omega(0)}^{'\mu\nu},
\end{equation}
where $T_{\Omega(0)}^{'\mu\nu}$ is the energy-momentum tensor of the $\Omega(0)$ substratum in the Fock coordinate system $S_{F}$.

We introduce the following notation in $S_{F}$
\begin{eqnarray}
&&\Theta^{'\mu\nu}=\partial'_{\lambda}\partial^{'\lambda}\psi^{'\mu\nu}-(\partial'_{\lambda}\partial^{'\mu}\psi^{'\nu\lambda}
+\partial'_{\lambda}\partial^{'\nu}\psi^{'\mu\lambda})\nonumber\\
&&+(\partial^{'\mu}\partial^{'\nu}\psi'+\eta^{'\mu\nu}\partial'_{\sigma}\partial'_{\lambda}\psi^{'\sigma\lambda})
-\eta^{'\mu\nu}\partial'_{\lambda}\partial^{'\lambda}\psi'.\ \ \ \ \ \label{Theta 870-1300}
\end{eqnarray}

Following Ref. \cite{WangXS201908}, we introduce the following definition of the contravariant energy-momentum tensor $T_{\omega}^{'\mu\nu}$ of vacuum in $S_{F}$
\begin{equation}\label{relationship 870-1400}
T_{\omega}^{'\mu\nu}=\frac{1}{f_{0}}\Psi^{'\mu\nu}+\frac{1}{f_{0}}\Theta^{'\mu\nu}+T_{\Omega(0)}^{'\mu\nu}.
\end{equation}

Using these definitions, we have the following result.
\begin{wcorollary_my}\label{field 870-1500}
The field equations (\ref{field 870-1010}) can be written as
\begin{eqnarray}
&&\partial'_{\lambda}\partial^{'\lambda}\psi^{'\mu\nu}-\partial'_{\lambda}\partial^{'\mu}\psi^{'\nu\lambda}
-\partial'_{\lambda}\partial^{'\nu}\psi^{'\mu\lambda}\nonumber\\
&&+\partial^{'\mu}\partial^{'\nu}\psi'+\eta^{'\mu\nu}\partial'_{\sigma}\partial'_{\lambda}\psi^{'\sigma\lambda}\nonumber\\
&&-\eta^{'\mu\nu}\partial'_{\lambda}\partial^{'\lambda}\psi'=-f_{0}(T^{'\mu\nu}-T_{\omega}^{'\mu\nu}).\label{field 870-1510}
\end{eqnarray}
\end{wcorollary_my}

We can verify that the field equations (\ref{field 870-1510}) are invariant under the following gauge transformation
\begin{equation}\label{gauge 870-1560}
\psi^{'\mu\nu} \rightarrow \psi^{'\mu\nu} + \partial^{'\mu}\Lambda^{\nu} +\partial^{'\nu}\Lambda^{\mu},
\end{equation}
where $\Lambda^{\mu}$ is an arbitrary vector field.

We introduce the following definition
\begin{equation}\label{definition 870-1600}
\phi^{'\mu\nu}=\psi^{'\mu\nu}-\frac{1}{2}\eta^{'\mu\nu}\psi'.
\end{equation}

Using Eqs.\ (\ref{definition 870-1600}), the field equations (\ref{field 870-1510}) can be written as
\begin{eqnarray}
&&\partial'_{\lambda}\partial^{'\lambda}\phi^{'\mu\nu}-\partial'_{\lambda}\partial^{'\mu}\phi^{'\nu\lambda}
-\partial'_{\lambda}\partial^{'\nu}\phi^{\mu\lambda}\nonumber\\
&&+\eta^{'\mu\nu}\partial'_{\sigma}\partial'_{\lambda}\phi^{'\sigma\lambda}= -f_{0}(T^{'\mu\nu}-T_{\omega}^{'\mu\nu}).\label{field 870-1700}
\end{eqnarray}

We introduce the following Hilbert gauge condition \cite{ThirringW1961}
\begin{equation}\label{Hilbert 870-1800}
\partial'_{\mu}\left ( \psi^{'\mu\nu}-\frac{1}{2}\eta^{'\mu\nu}\psi' \right )=0.
\end{equation}

Using Eqs.\ (\ref{definition 870-1600}), the Hilbert gauge condition Eqs.\ (\ref{Hilbert 870-1800}) simplifies to
\begin{equation}\label{Hilbert 870-1900}
\partial'_{\mu}\phi^{'\mu\nu}=0.
\end{equation}

If we impose the Hilbert gauge condition Eqs.\ (\ref{Hilbert 870-1900}) on the fields, then the field equations  (\ref{field 870-1700}) simplify to
\begin{equation}\label{field 870-2000}
\partial'_{\lambda}\partial^{'\lambda}\phi^{'\mu\nu}=-f_{0}(T^{'\mu\nu}-T_{\omega}^{'\mu\nu}).
\end{equation}

The field equations (\ref{field 870-2000}) can also be written as
\begin{equation}\label{field 870-2100}
\eta^{'\alpha\beta}\frac{\partial^{2}\phi^{'\mu\nu}}{\partial x^{'\alpha} \partial x^{'\beta}}=-f_{0}(T^{'\mu\nu}-T_{\omega}^{'\mu\nu}).
\end{equation}

\section{Generalized Einstein's equations in a special class of non-inertial reference frames \label{sec 900}}
\begin{definition_my}\label{Einstein 900-2800}
The Einstein tensor $G_{\mu\nu}$ is defined by
\begin{equation}\label{Einstein 900-2810}
G_{\mu\nu} \equiv R_{\mu\nu}-\frac{1}{2}g_{\mu\nu}R,
\end{equation}
where $g_{\mu\nu}$ is a metric tensor of a Riemannian spacetime,
$R_{\mu\nu}$ is the Ricci tensor, $R \equiv g^{\mu\nu}R_{\mu\nu}$, $g^{\mu\nu}$ is the corresponding contravariant tensor of $g_{\mu\nu}$ such that $g_{\mu\lambda}g^{\lambda\nu}=\delta_{\mu}^{\nu}$ (\cite{LiuL2004}, p.\ 40).
\end{definition_my}

Similar to Ref. \cite{WangXS201908}, we introduce the following definition of a metric tensor $g_{\mu\nu}$ of a Riemannian spacetime.
\begin{definition_my}\label{metric 900-50}
\begin{equation}\label{metric 900-55}
\tilde{g}^{\mu\nu} \equiv \sqrt{-g_{0}}g^{\mu\nu} \equiv \eta^{'\mu\nu}-2f_{0}\phi^{'\mu\nu},
\end{equation}
where $g_{0} = \mathrm{Det} \ g_{\mu\nu}$.
\end{definition_my}

Applying similar methods of V. Fock (\cite{FockV1964}, p.\ 422-430), Fock's theorem of the Einstein tensor $G_{\mu\nu}$ in the Galilean coordinates (\cite{FockV1964}, p.\ 429) can be generalized to non-inertial coordinate systems $(x'^{0}, \ x'^{1},  \  x'^{2},  \  x'^{3})$.
\begin{proposition_my}\label{Fock 870-2100}
The contravariant Einstein tensor $G^{\mu\nu}$ in a non-inertial coordinate systems $(x'^{0}, \ x'^{1},  \  x'^{2},  \  x'^{3})$ can be written as
\begin{eqnarray}
G^{\mu\nu}&=&\frac{1}{2g_{0}}\tilde{g}^{\alpha\beta}\frac{\partial^{2} \tilde{g}^{\mu\nu}}{\partial x^{'\alpha} \partial x^{'\beta}}
+\Pi^{'\mu,\alpha\beta}\Pi_{\alpha\beta}^{'\nu}-\frac{1}{2}y^{'\mu}y^{'\nu} \nonumber\\
&&+\frac{1}{2}g^{\mu\nu}(L'+B')-B^{'\mu\nu},\label{Fock 870-2110}
\end{eqnarray}
where
\begin{equation}\label{Fock 870-2120}
\Pi^{'\mu,\alpha\beta}\equiv\frac{1}{2g_{0}}
\left ( \tilde{g}^{\alpha\lambda}\frac{\partial \tilde{g}^{\mu\beta}}{\partial x^{'\lambda}}
+\tilde{g}^{\beta\lambda}\frac{\partial \tilde{g}^{\mu\alpha}}{\partial x^{'\lambda}}-\tilde{g}^{\mu\lambda}\frac{\partial \tilde{g}^{\alpha\beta}}{\partial x^{'\lambda}}\right ),
\end{equation}
\begin{equation}\label{Fock 870-2130}
\Pi_{\alpha\beta}^{'\nu}\equiv g_{\alpha\lambda}g_{\beta\sigma}\Pi^{'\nu,\lambda\sigma},
\end{equation}
\begin{equation}\label{Fock 870-2140}
\Gamma^{'\alpha}\equiv g^{\sigma\lambda}\Gamma_{\sigma\lambda}^{'\alpha},
\end{equation}
\begin{equation}\label{Christoffel 870-2145}
\Gamma_{\alpha\beta}^{'\nu}\equiv \frac{1}{2}g^{\mu\nu}
\left ( \frac{\partial g_{\mu\alpha}}{\partial x^{'\beta}}
+\frac{\partial g_{\mu\beta}}{\partial x^{'\alpha}}-\frac{\partial g_{\alpha\beta}}{\partial x^{'\mu}}\right )
\end{equation}
\begin{equation}\label{Fock 870-2150}
\Gamma^{'\mu\nu} \equiv  \frac{1}{2}
\left ( g^{\mu\alpha}\frac{\partial \Gamma^{'\nu}}{\partial x^{'\alpha}}
 + g^{\nu\alpha}\frac{\partial \Gamma^{'\mu}}{\partial x^{'\alpha}}
- \frac{\partial g^{\mu\nu}}{\partial x^{'\alpha}}\Gamma^{'\alpha}\right ),
\end{equation}
\begin{equation}\label{Fock 870-2160}
y'_{\beta}\equiv \frac{\partial(\lg\sqrt{-g_{0}})}{\partial x^{'\beta}}, \ y^{'\alpha}\equiv g^{\alpha\beta}y'_{\beta},
\end{equation}
\begin{equation}\label{Fock 870-2170}
L' \equiv  -\frac{1}{2\sqrt{-g_{0}}}\Gamma_{\alpha\beta}^{'\nu}\frac{\partial  \tilde{g}^{\alpha\beta}}{\partial x^{'\nu}}+\frac{1}{2}y'_{\nu}y'^{\nu},
\end{equation}
\begin{equation}\label{Fock 870-2180}
B^{'\mu\nu}\equiv \Gamma^{'\mu\nu}+\frac{1}{2}(y^{'\mu}\Gamma^{'\nu}+y^{'\nu}\Gamma^{'\mu}),\ B'\equiv g_{\mu\nu}B^{'\mu\nu}.
\end{equation}
\end{proposition_my}

A proof of Proposition \ref{Fock 870-2100} can be found in the Appendix.

\begin{theorem_my}\label{field 900-100}
In a Fock coordinate system $S_{F}$, we have the following field equations
\begin{eqnarray}
&&G^{\mu\nu}-\frac{1}{2g_{0}}\left (\sqrt{-g_{0}}g^{\alpha\beta}-\eta^{'\alpha\beta}\right )\frac{\partial^{2}(\sqrt{-g_{0}}g^{\mu\nu})}{\partial x^{'\alpha} \partial x^{'\beta}} \nonumber\\
&&-\frac{1}{2g_{0}}\eta^{'\alpha\beta}\frac{\partial^{2} \eta^{'\mu\nu}}{\partial x^{'\alpha} \partial x^{'\beta}}-\Pi^{'\mu,\alpha\beta}\Pi_{\alpha\beta}^{'\nu} + \frac{1}{2}y^{'\mu}y^{'\nu}\nonumber\\
&&-\frac{1}{2}g^{\mu\nu}(L'+B')+B^{'\mu\nu}=\frac{f_{0}^{2}}{g_{0}}(T^{'\mu\nu}-T^{'\mu\nu}_{\omega}).
\ \ \ \ \ \label{Einstein 900-120}
\end{eqnarray}
\end{theorem_my}
{\bfseries{Proof of Theorem \ref{field 900-100}.}} Using Eqs.\ (\ref{metric 900-55}), Eqs.\ (\ref{Fock 870-2110}) can be written as
\begin{eqnarray}
&&G^{\mu\nu}=\frac{1}{2g_{0}}\left (\sqrt{-g_{0}}g^{\alpha\beta}-\eta^{'\alpha\beta}+\eta^{'\alpha\beta}\right ) \nonumber\\
&&\times\frac{\partial^{2}(\eta^{'\mu\nu}-2f_{0}\phi^{'\mu\nu})}{\partial x^{'\alpha} \partial x^{'\beta}}+\Pi^{'\mu,\alpha\beta}\Pi_{\alpha\beta}^{'\nu}\nonumber\\
&&-\frac{1}{2}y^{'\mu}y^{'\nu}+\frac{1}{2}g^{\mu\nu}(L'+B')-B^{'\mu\nu}\nonumber\\
&&=\frac{1}{2g_{0}}\left (\sqrt{-g_{0}}g^{\alpha\beta}-\eta^{'\alpha\beta}\right )\frac{\partial^{2}(\eta^{'\mu\nu}-2f_{0}\phi^{'\mu\nu})}{\partial x^{'\alpha} \partial x^{'\beta}} \nonumber\\
&&+\frac{1}{2g_{0}}\eta^{'\alpha\beta}\frac{\partial^{2}\eta^{'\mu\nu}}{\partial x^{'\alpha} \partial x^{'\beta}}-\frac{f_{0}}{g_{0}}\eta^{'\alpha\beta}\frac{\partial^{2}\phi^{'\mu\nu}}{\partial x^{'\alpha} \partial x^{'\beta}} \nonumber\\
&&+\Pi^{'\mu,\alpha\beta}\Pi_{\alpha\beta}^{'\nu}
-\frac{1}{2}y^{'\mu}y^{'\nu}\nonumber\\
&&+\frac{1}{2}g^{\mu\nu}(L'+B')-B^{'\mu\nu}.\label{Einstein 900-130}
\end{eqnarray}

Using Eqs.\ (\ref{metric 900-55}) and Eqs.\ (\ref{field 870-2100}), Eqs.\ (\ref{Einstein 900-130}) can be written as
Eqs.\ (\ref{Einstein 900-120}). $\Box$

We need to study the relationships between Eqs.\ (\ref{Einstein 900-120}) and the Einstein's field equations.
In a harmonic coordinates system, we have (\cite{FockV1964}, p.\ 254)
\begin{equation}\label{harmonic 900-150}
\Gamma^{'\nu} = \Gamma^{'\mu\nu} = B^{'\mu\nu} = B' = 0.
\end{equation}

Using Eqs.\ (\ref{harmonic 900-150}) and Eqs.\ (\ref{Einstein 900-120}), we have the following result.
\begin{wcorollary_my}\label{field 900-200}
In a Fock coordinate system $S_{F}$ the field equations (\ref{Einstein 900-120}) can be written as
\begin{eqnarray}
G^{\mu\nu}&-&\frac{1}{2g_{0}}\left (\sqrt{-g_{0}}g^{\alpha\beta}-\eta^{'\alpha\beta}\right )\frac{\partial^{2}(\sqrt{-g_{0}}g^{\mu\nu})}{\partial x^{'\alpha} \partial x^{'\beta}} \nonumber\\
&-&\frac{1}{2g_{0}}\eta^{'\alpha\beta}\frac{\partial^{2} \eta^{'\mu\nu}}{\partial x^{'\alpha} \partial x^{'\beta}}-\Pi^{'\mu,\alpha\beta}\Pi_{\alpha\beta}^{'\nu}\nonumber\\
&+& \frac{1}{2}y^{'\mu}y^{'\nu}-\frac{1}{2}g^{\mu\nu}L'\nonumber\\
&=&\frac{f_{0}^{2}}{g_{0}}(T^{'\mu\nu}-T^{'\mu\nu}_{\omega}).\label{Einstein 900-210}
\end{eqnarray}
\end{wcorollary_my}

\begin{definition_my}\label{weak 900-300}
If the following conditions are valid
\begin{eqnarray}
&&\eta^{'\mu\nu} \approx \eta^{\mu\nu}, \label{quasiinertial 900-310} \\
&&\left | \frac{1}{2}\eta^{'\alpha\beta}\frac{\partial^{2} \eta^{'\mu\nu}}{\partial x^{'\alpha} \partial x^{'\beta}} \right | \ll \left | f_{0}^{2}(T^{'\mu\nu}-T^{'\mu\nu}_{\omega}) \right |, \ \ \ \ \label{quasiinertial 900-320}
\end{eqnarray}
then we call this reference frame quasi-inertial.
\end{definition_my}

Using Eqs.\ (\ref{quasiinertial 900-320}) and Eqs.\ (\ref{Einstein 900-210}), we have the following result.
\begin{wcorollary_my}\label{field 900-400}
If the reference frame $S_{F}$ is quasi-inertial, then, the field equations (\ref{Einstein 900-210}) can be written as
\begin{eqnarray}
&&G^{\mu\nu}-\frac{1}{2g_{0}}\left (\sqrt{-g_{0}}g^{\alpha\beta}-\eta^{'\alpha\beta}\right )\frac{\partial^{2}(\sqrt{-g_{0}}g^{\mu\nu})}{\partial x^{'\alpha} \partial x^{'\beta}} \nonumber\\
&&-\Pi^{'\mu,\alpha\beta}\Pi_{\alpha\beta}^{'\nu} + \frac{1}{2}y^{'\mu}y^{'\nu}-\frac{1}{2}g^{\mu\nu}L'\nonumber\\
&&\approx \frac{f_{0}^{2}}{g_{0}}(T^{'\mu\nu}-T^{'\mu\nu}_{\omega}).\label{Einstein 900-410}
\end{eqnarray}
\end{wcorollary_my}

Eqs.\ (\ref{Einstein 900-410}) are only valid approximately in a quasi-inertial Fock coordinate system $S_{F}$. Now we consider weak fields.

\begin{definition_my}\label{weak 900-330}
If $\phi^{'\mu\nu}$ and their first and higher derivatives satisfy the following conditions
\begin{eqnarray}
&& \left | 2f_{0}\phi^{'\mu\nu} \right |\ll 1, \label{weak 900-336} \\
&&\left | \frac{\partial^{j+k} (2f_{0}\phi^{'\mu\nu})}{\partial (x^{'\alpha})^{j}\partial (x^{'\beta})^{k}} \right | \ll 1, j+k=1, 2, 3, \cdots \ \ \ \label{weak 900-338}
\end{eqnarray}
then we call this field $\phi^{'\mu\nu}$ weak.
\end{definition_my}

Following similar ideas as in Ref. \cite{WangXS201908}, we introduce the following assumption.
\begin{assumption_my}\label{assumption 900-350}
For weak fields, the following relationships are valid
\begin{equation}\label{relationship 900-360}
T^{'\mu\nu}-T^{'\mu\nu}_{\omega} \approx T_{m}^{'\mu\nu}+T_{\Omega(1)}^{'\mu\nu}.
\end{equation}
\end{assumption_my}

Similar to Ref. \cite{WangXS201908}, we have the following result.
\begin{wcorollary_my}\label{field 900-500}
Suppose that (1) the Fock coordinate system $S_{F}$ is quasi-inertial; (2) the field is weak. Then, the field equations (\ref{Einstein 900-410}) reduce to
\begin{equation}\label{Einstein 900-510}
R_{\mu\nu}-\frac{1}{2}g_{\mu\nu}R \approx \frac{f_{0}^{2}}{g_{0}}\left ( T^{'m}_{\mu\nu}+T^{'\Omega(1)}_{\mu\nu}\right ).
\end{equation}
\end{wcorollary_my}
{\bfseries{Proof of Corollary \ref{field 900-500}.}}
According to Definition \ref{weak 900-330}, $f_{0}\phi^{'\mu\nu}$ and their first and higher derivatives are small quantities of order $\varepsilon$, where $|\varepsilon| \ll 1$ is a small quantity.
Since the reference frame is quasi-inertial, Eqs.\ (\ref{quasiinertial 900-310}) are valid. Using Eqs.\ (\ref{quasiinertial 900-310}),  Eqs.\ (\ref{metric 900-55}) can be written as
\begin{equation}\label{metric 900-550}
\sqrt{-g_{0}}g^{\mu\nu} \approx \eta^{\mu\nu}-2f_{0}\phi^{'\mu\nu}.
\end{equation}

Since the field is weak, Eqs.\ (\ref{weak 900-336}) and Eqs.\ (\ref{weak 900-338}) are valid. Thus, applying Eqs.\ (\ref{weak 900-336}) and Eqs.\ (\ref{metric 900-550}), we have the following estimations of the order of magnitude of the following quantities
\begin{equation}\label{order 900-700}
\sqrt{-g_{0}}g^{\alpha\beta}-\eta^{'\alpha\beta} \sim \varepsilon.
\end{equation}

Using Eqs.\ (\ref{weak 900-338}), we have the following estimations
\begin{equation}\label{order 900-800}
\frac{\partial g_{\mu\nu}}{\partial x^{'\alpha}}\sim \frac{\partial g^{\mu\nu}}{\partial x^{'\alpha}}\sim \varepsilon.
\end{equation}

Applying Eqs.\ (\ref{metric 900-550}) and Eqs.\ (\ref{weak 900-338}), we have the following estimations
\begin{equation}\label{order 900-900}
\frac{\partial^{2}(\sqrt{-g_{0}}g^{\mu\nu})}{\partial x^{'\alpha} \partial x^{'\beta}} = \frac{\partial^{2}(-2f_{0}\phi^{'\mu\nu})}{\partial x^{'\alpha} \partial x^{'\beta}} \sim \varepsilon.
\end{equation}

Thus, using Eqs.\ (\ref{order 900-700}) and Eqs.\ (\ref{order 900-900}), we have the following estimations
\begin{equation}\label{order 900-1000}
\left (\sqrt{-g_{0}}g^{\alpha\beta}-\eta^{'\alpha\beta}\right )\frac{\partial^{2}(\sqrt{-g_{0}}g^{\mu\nu})}{\partial x^{'\alpha} \partial x^{'\beta}} \sim \varepsilon^{2}.
\end{equation}

Applying Eqs.\ (\ref{Fock 870-2120}), Eqs.\ (\ref{Fock 870-2130}) and Eqs.\ (\ref{order 900-800}), we have the following estimations
\begin{equation}\label{order 900-1200}
\Pi^{'\mu,\alpha\beta} \sim \Pi_{\alpha\beta}^{'\nu} \sim \varepsilon.
\end{equation}

Using Eqs.\ (\ref{Fock 870-2160}), we have the following relationship (\cite{FockV1964}, p.\ 143)
\begin{equation}\label{formula 900-1300}
y'_{\beta} = \Gamma_{\beta\nu}^{'\nu}.
\end{equation}

We also have (\cite{FockV1964}, p.\ 143)
\begin{equation}\label{formula 900-1400}
\Gamma_{\beta\nu}^{'\nu} = \frac{1}{2}g^{\mu\nu}\frac{\partial g_{\mu\nu}}{\partial x^{'\beta}}.
\end{equation}

Applying Eqs.\ (\ref{formula 900-1300}), Eqs.\ (\ref{formula 900-1400}) and Eqs.\ (\ref{order 900-800}),  we have the following estimations
\begin{equation}\label{order 900-1500}
y'_{\beta} \sim \varepsilon.
\end{equation}

Using Eqs.\ (\ref{Fock 870-2160}) and Eqs.\ (\ref{order 900-1500}),  we have the following estimations
\begin{equation}\label{order 900-1600}
y^{'\alpha} \sim \varepsilon.
\end{equation}

Similar to the case of the Galilean coordinates, we have (\cite{FockV1964}, p.\ 430)
\begin{equation}\label{L 900-1650}
L' = -\frac{1}{2}\Gamma_{\alpha\beta}^{'\nu}\frac{\partial g^{\alpha\beta}}{\partial x^{'\nu}}-\Gamma^{'\alpha}y'_{\alpha}.
\end{equation}

Applying Eqs.\ (\ref{harmonic 900-150}),  Eq.\ (\ref{L 900-1650}) can be written as
\begin{equation}\label{relation 900-1700}
L' = -\frac{1}{2}\Gamma_{\alpha\beta}^{'\nu}\frac{\partial g^{\alpha\beta}}{\partial x^{'\nu}}.
\end{equation}

Using Eq.\ (\ref{relation 900-1700}), Eqs.\ (\ref{Christoffel 870-2145}) and Eqs.\ (\ref{order 900-800}),  we have the following estimation
\begin{equation}\label{order 900-1800}
L' \sim \varepsilon^{2}.
\end{equation}

Applying Eqs.\ (\ref{order 900-1000}), Eqs.\ (\ref{order 900-1200}), Eqs.\ (\ref{order 900-1600}) and Eq.\ (\ref{order 900-1800}), we see that the second to the fifth term on the left side of Eqs.\ (\ref{Einstein 900-410}) are all small quantities of order $\varepsilon^{2}$. Ignoring all these small quantities of order $\varepsilon^{2}$ in Eqs.\ (\ref{Einstein 900-410}) and using Eqs.\ (\ref{relationship 900-360}), we obtain
\begin{equation}\label{Einstein 900-1900}
G^{\mu\nu} \approx \frac{f_{0}^{2}}{g_{0}}\left ( T_{m}^{'\mu\nu}+T_{\Omega(1)}^{'\mu\nu}\right ).
\end{equation}

Applying the rules of lowering the indexes of tensors, i.e., $G^{\mu\nu}=g^{\mu\sigma}g^{\nu\lambda}G_{\sigma\lambda}$, $T_{m}^{'\mu\nu}=g^{\mu\sigma}g^{\nu\lambda}T^{'m}_{\sigma\lambda}$, $T_{\Omega(1)}^{'\mu\nu}=g^{\mu\sigma}g^{\nu\lambda}T^{'\Omega(1)}_{\sigma\lambda}$, Eqs.\ (\ref{Einstein 900-1900}) can be written as
\begin{equation}\label{Einstein 900-2000}
G_{\sigma\lambda} \approx \frac{f_{0}^{2}}{g_{0}}\left ( T^{'m}_{\sigma\lambda}+T^{'\Omega(1)}_{\sigma\lambda}\right ).
\end{equation}

Putting Eqs.\ (\ref{Einstein 900-2810}) into Eqs.\ (\ref{Einstein 900-2000}), we obtain Eqs.\ (\ref{Einstein 900-510}). $\Box$

Using Eq.\ (\ref{f0 850-950}), the field equations (\ref{Einstein 900-510}) can be written as
\begin{equation}\label{Einstein 900-2100}
R_{\mu\nu}-\frac{1}{2}g_{\mu\nu}R \approx \frac{1}{g_{0}}\frac{8\pi \gamma_{N}}{c^{4}}\left ( T^{'m}_{\mu\nu}+T^{'\Omega(1)}_{\mu\nu}\right ).
\end{equation}

Similar to Ref. \cite{WangXS201908}, we have the following result.
\begin{wcorollary_my}\label{field 900-600}
Suppose that (1) the Fock coordinate system $S_{F}$ is quasi-inertial; (2) the field is weak; (3) $g_{0} \approx -1$. Then, the field equations (\ref{Einstein 900-510}) reduce to
\begin{equation}\label{field 900-610}
R_{\mu\nu}-\frac{1}{2}g_{\mu\nu}R \approx -f_{0}^{2} \left ( T^{'m}_{\mu\nu}+T^{'\Omega(1)}_{\mu\nu}\right ).
\end{equation}
\end{wcorollary_my}

We introduce the following notation
\begin{equation}\label{kappa 900-2200}
\kappa = f_{0}^{2}=\frac{8\pi \gamma_{N}}{c^{4}}.
\end{equation}

Using Eq.\ (\ref{kappa 900-2200}), the field equations (\ref{Einstein 900-510}) can be written as
\begin{equation}\label{Einstein 900-2300}
R_{\mu\nu}-\frac{1}{2}g_{\mu\nu}R \approx \frac{\kappa}{g_{0}}\left ( T^{'m}_{\mu\nu}+T^{'\Omega(1)}_{\mu\nu}\right ).
\end{equation}

Using Eq.\ (\ref{kappa 900-2200}), the field equations (\ref{field 900-610}) can be written as
\begin{equation}\label{field 900-2400}
R_{\mu\nu}-\frac{1}{2}g_{\mu\nu}R \approx -\kappa \left ( T^{'m}_{\mu\nu}+T^{'\Omega(1)}_{\mu\nu}\right ).
\end{equation}

We notice that the field equations (\ref{field 900-2400}) are the Einstein's equations \cite{MollerC1972,WeinbergS1972,MisnerC1973} if we suppose that $T^{'\Omega(1)}_{\mu\nu}\approx 0$.  Therefore, the field equations (\ref{Einstein 900-120}) are generalizations of the Einstein's equations in some special non-inertial reference frames. Thus, all known experiments of gravitational phenomena which support the theory of general relativity may also be explained by this
theory of gravity based on the theory of vacuum mechanics \cite{WangXS200804,WangXS200810,WangXS2014PhysicsEssays,WangXS201908}.

\section{Discussion \label{sec 1000}}
Although the field equations (\ref{Einstein 900-120}) are generalizations of the Einstein's equations, there exist at least the following differences between this theory and the theory of general relativity.

1. In the theory of general relativity, the Einstein's equations are assumptions \cite{MollerC1972,WeinbergS1972,MisnerC1973}. Although A. Einstein introduced his new concept of gravitational aether (\cite{Kostro2000}, p.\ 63-113), he did not derive his equations theoretically based on his new concept of the gravitational aether. In our theory, the generalized Einstein's equations (\ref{Einstein 900-120}) are derived by methods of special relativistic continuum mechanics based on some assumptions.

2. Although the theory of general relativity is a field theory of gravity, the definitions of gravitational fields are not based on continuum mechanics \cite{MollerC1972,WeinbergS1972,MisnerC1973,Eringen1980}.
Because of the absence of a continuum, the theory of general relativity may be regarded as a phenomenological theory of gravity. In our theory, gravity is transmitted by the $\Omega (0)$ substratum. The tensorial potential $\psi'_{\mu\nu}$ of gravitational fields are defined based on special relativistic continuum mechanics.

3. In Einstein's theory, the concept of Riemannian spacetime is introduced together with the field equations \cite{MollerC1972,WeinbergS1972,MisnerC1973}. The theory of general relativity cannot provide a physical definition of the metric tensor of the Riemannian spacetime. In our theory, the background spacetime is the Minkowski spacetime. However, the initial flat background spacetime is no longer physically observable. According to the equation of motion of a point particle in gravitational field in inertial reference frames \cite{WangXS201908},  to the first order of $f_{0}\psi_{\mu\nu}$, the physically observable spacetime is a Riemannian spacetime with the metric tensor $g_{\mu\nu}$. The metric tensor $g_{\mu\nu}$ is defined based on the tensorial potential $\psi'_{\mu\nu}$ of gravitational fields.

4. The masses of particles are constants in the theory of general relativity \cite{MollerC1972,WeinbergS1972,MisnerC1973}. In our theory, the masses of particles are functions of time $t$ \cite{WangXS200810}.

5. The gravitational constant $\gamma_{N}$ is a constant in the theory of general relativity \cite{MollerC1972,WeinbergS1972,MisnerC1973}. The theory of general relativity cannot provide a derivation of $\gamma_{N}$. In our theory, the parameter $\gamma_{N}$ is derived theoretically. From Eq.\ (\ref{constant 820-210}), we see that $\gamma_{N}$ depends on time $t$.

6. In our theory, the parameter $\gamma_{N}$ in Eq.\ (\ref{constant 820-210}) depends on the density $\rho_{0}$ of the $\Omega (0)$ substratum. If $\rho_{0}$ varies from place to place, i.e., $\rho_{0}=\rho_{0}(t, x, y, z)$, then the space dependence of the gravitational constant $\gamma_{N}$ can be seen from Eq.\ (\ref{constant 820-210}).

7. The Einstein's equations are supposed to be valid in all reference frames \cite{MollerC1972,WeinbergS1972,MisnerC1973}. It is known that general relativity is a pure theory and contains no adjustable constants. So the predictions of general relativity are fixed. Thus, every experimental test of the theory is important. For more than 100 years, experimental tests of general relativity are carried out only in the solar system \cite{WillCM1993,WillCW2014}. However, the solar system can be regarded approximately as an inertial reference frame. Therefore, it is still not clear whether the Einstein's equations are valid in all non-inertial reference frames or not. However, in our theory the generalized Einstein's equations (\ref{Einstein 900-120}) are valid only in Fock coordinate systems. Whether it is possible for us to derive generalized Einstein's equations in other non-inertial reference frames is an interesting question. V. Fock said (\cite{FockV1964}, p.\ 394):"{\itshape  But physically the 'general principle of relativity', in the sense that corresponding processes exist in arbitrary reference frames, does not hold at all. Therefore Einstein's conclusion that all reference frames are physically equivalent, is without foundation.}"

8. The Einstein's equations are rigorous \cite{MollerC1972,WeinbergS1972,MisnerC1973}. However, in our theory, Eqs.\ (\ref{field 900-2400}) are valid approximately under some assumptions.

It is interesting whether it is possible for us to detect some of these differences by experiments.

\section{Conclusion \label{sec 1100}}
Inspired by the mathematical forms of the dynamical gravitational potentials in inertial reference frames, we establish mathematical models of inertial potential and inertial force Lagrangian in non-inertial reference frames. Field equations of gravitation in Fock coordinate systems are derived based on the Euler-Lagrange equations. Applying the generalized Fock's theorem, generalized Einstein's equations in Fock coordinate systems are derived. If the field is weak and the reference frame is quasi-inertial, these generalized Einstein's equations reduce to Einstein's equations.

\section*{Acknowledgments \label{sec 1200}}
This work was partly supported by the Doctor Research Foundation of Henan Polytechnic University (Grant No. B2012-069). I would like to thank an anonymous reviewer for his important remarks and opinions.

\section*{Appendix}\label{sec 1300}
{\bfseries{Proof of Proposition \ref{Fock 870-2100}.}} The definition of the covariant second rank curvature tensor $R_{\mu\nu}$ is (\cite{FockV1964}, p.\ 422)
\begin{equation}\label{curvature 1300-100}
R_{\mu\nu}\equiv g^{\alpha\beta}R_{\mu\alpha,\beta\nu},
\end{equation}
where
\begin{eqnarray}
R_{\mu\alpha,\beta\nu}&\equiv &\frac{1}{2}
\left ( \frac{\partial^{2} g_{\mu\nu}}{\partial x^{'\alpha}\partial x^{'\beta}}
+\frac{\partial^{2} g_{\alpha\beta}}{\partial x^{'\mu}\partial x^{'\nu}}-\frac{\partial^{2} g_{\nu\alpha}}{\partial x^{'\mu}\partial x^{'\beta}} \right. \nonumber\\
&& \left. -\frac{\partial^{2} g_{\mu\beta}}{\partial x^{'\nu}\partial x^{'\alpha}}\right )
-g_{\sigma\lambda}\Gamma_{\mu\beta}^{'\sigma}\Gamma_{\nu\alpha}^{'\lambda}\nonumber\\
&&+g_{\sigma\lambda}\Gamma_{\mu\nu}^{'\sigma}\Gamma_{\alpha\beta}^{'\lambda}, \label{curvature 1300-200}
\end{eqnarray}
is the fourth rank curvature tensor.

The contravariant curvature tensor $R^{\mu\nu}$ can be obtained by raising the indices (\cite{FockV1964}, p.\ 156)
\begin{equation}\label{curvature 1300-300}
R^{\mu\nu}=g^{\mu\sigma}g^{\nu\lambda}R_{\sigma\lambda}.
\end{equation}

Following similar methods of V. Fock (\cite{FockV1964}, p.\ 425), we have
\begin{equation}\label{curvature 1300-400}
R^{\mu\nu}=\frac{1}{2}g_{\alpha\beta}\frac{\partial^{2} g^{\mu\nu}}{\partial x^{'\alpha}\partial x^{'\beta}}
-\Gamma^{'\mu\nu}+\Gamma^{'\mu,\alpha\beta}\Gamma_{\alpha\beta}^{'\nu}.
\end{equation}

The definition of the invariant of the curvature tensor is (\cite{FockV1964}, p.\ 425)
\begin{equation}\label{invariant 1300-500}
R\equiv g_{\mu\nu}R^{\mu\nu}.
\end{equation}

Following similar methods of V. Fock (\cite{FockV1964}, p.\ 428), we have
\begin{equation}\label{invariant 1300-600}
R=g^{\alpha\beta}y'_{\alpha\beta}-\Gamma^{'\alpha}y'_{\alpha}-\Gamma'-L',
\end{equation}
where
\begin{equation}\label{y 1300-700}
y'_{\alpha\beta}\equiv \frac{\partial^{2}\lg\sqrt{-g_{0}}}{\partial x^{'\alpha}\partial x^{'\beta}},
\end{equation}
\begin{equation}\label{Gamma 1300-800}
\Gamma'\equiv g_{\mu\nu}\Gamma^{'\mu\nu},
\end{equation}
\begin{equation}\label{L 1300-850}
L' \equiv  -\frac{1}{2}\Gamma_{\alpha\beta}^{'\nu}\frac{\partial g^{\alpha\beta}}{\partial x^{'\nu}}-\Gamma^{'\alpha}\frac{\partial(\lg\sqrt{-g_{0}})}{\partial x^{'\alpha}},
\end{equation}
The second derivative of $\tilde{g}^{\mu\nu}$ is (\cite{FockV1964}, p.\ 428)
\begin{eqnarray}
\frac{\partial^{2} \tilde{g}^{\mu\nu}}{\partial x^{'\alpha}\partial x^{'\beta}}=&&\sqrt{-g_{0}}
\left ( \frac{\partial^{2} g^{\mu\nu}}{\partial x^{'\alpha}\partial x^{'\beta}}
+y'_{\beta}\frac{\partial g_{\mu\nu}}{\partial x^{'\alpha}}+y'_{\alpha}\frac{\partial g_{\mu\nu}}{\partial x^{'\beta}} \right. \nonumber\\
&&\left. +y'_{\alpha\beta}g^{\mu\nu} + y'_{\alpha}y'_{\beta}g^{\mu\nu}\right ). \label{derivative 1300-900}
\end{eqnarray}

Multiplying $g^{\alpha\beta}$, Eqs.\ (\ref{derivative 1300-900}) can be written as (\cite{FockV1964}, p.\ 428)
\begin{eqnarray}
g^{\alpha\beta}\frac{\partial^{2} \tilde{g}^{\mu\nu}}{\partial x^{'\alpha}\partial x^{'\beta}}=&&\sqrt{-g_{0}}
\left ( \frac{\partial^{2} g^{\mu\nu}}{\partial x^{'\alpha}\partial x^{'\beta}}
+2y'^{\alpha}\frac{\partial g^{\mu\nu}}{\partial x^{'\alpha}}\right. \nonumber\\
&&\left. +g^{\mu\nu}g^{\alpha\beta}y'_{\alpha\beta} + g^{\mu\nu}y'_{\alpha}y'^{\alpha}\right ). \ \ \ \ \ \ \ \label{derivative 1300-1000}
\end{eqnarray}

Using Eqs.\ (\ref{curvature 1300-400}) and Eqs.\ (\ref{invariant 1300-600}), we have (\cite{FockV1964}, p.\ 428)
\begin{eqnarray}
R_{\mu\nu}-\frac{1}{2}g_{\mu\nu}R&=&\frac{1}{2}
\left ( g^{\alpha\beta}\frac{\partial^{2} g_{\mu\nu}}{\partial x^{'\alpha}\partial x^{'\beta}}
+g^{\mu\nu}g^{\alpha\beta}y'_{\alpha\beta}\right ) \nonumber\\
&&+\frac{1}{2}g^{\mu\nu}(\Gamma^{'\alpha}y'_{\alpha}+\Gamma'+L')\nonumber\\
&&-\Gamma^{'\mu\nu}+\Gamma^{'\mu,\alpha\beta}\Gamma_{\alpha\beta}^{'\nu}. \label{curvature 1300-1100}
\end{eqnarray}

Comparing Eqs.\ (\ref{curvature 1300-1100}) and Eqs.\ (\ref{derivative 1300-1000}), we have (\cite{FockV1964}, p.\ 429)
\begin{eqnarray}
R_{\mu\nu}-\frac{1}{2}g_{\mu\nu}R&=&\frac{1}{2\sqrt{-g_{0}}}g^{\alpha\beta}\frac{\partial^{2} \tilde{g}_{\mu\nu}}{\partial x^{'\alpha}\partial x^{'\beta}}\nonumber\\
&&+\frac{1}{2}g^{\mu\nu}(y'_{\alpha}y'^{\alpha}+\Gamma^{'\alpha}y'_{\alpha}+\Gamma'+L')\nonumber\\
&&-\Gamma^{'\mu\nu}+y'^{\alpha}\frac{\partial g^{\mu\nu}}{\partial x^{'\alpha}}+\Gamma^{'\mu,\alpha\beta}\Gamma_{\alpha\beta}^{'\nu}.\ \ \ \ \ \ \ \label{curvature 1300-1200}
\end{eqnarray}

Using the notations defined in Eqs.\ (\ref{Einstein 900-2810}), Eqs.\ (\ref{Fock 870-2120}), Eqs.\ (\ref{Fock 870-2130}) and Eqs.\ (\ref{Fock 870-2180}), Eqs.\ (\ref{curvature 1300-1200}) can also be written as Eqs.\ (\ref{Fock 870-2110}) (\cite{FockV1964}, p.\ 429-430). $\Box$

\end{document}